\documentclass[aps,pre,twocolumn,showpacs,superscriptaddress,preprintnumbers,ams
math,amssymb]{revtex4}

\usepackage{amsmath}
\usepackage{graphicx}
\usepackage{dcolumn}
\usepackage{bm}
\usepackage{float}
\usepackage{color}
\usepackage{subfig}
\usepackage{textcomp}

\begin{document}

\author{Q. Moreno}\affiliation{University of Bordeaux, Centre Lasers Intenses et Applications, CNRS, CEA, UMR 5107, F-33405 Talence, France}

\author{M. E. Dieckmann}\affiliation{Department of Science and Technology, Link\"oping University, SE-60174 Norrk\"oping, Sweden}

\author{X. Ribeyre}\affiliation{University of Bordeaux, Centre Lasers Intenses et Applications, CNRS, CEA, UMR 5107, F-33405 Talence, France}

\author{E. d'Humi\`eres}\affiliation{University of Bordeaux, Centre Lasers Intenses et Applications, CNRS, CEA, UMR 5107, F-33405 Talence, France}

\date{\today}

\title{Quasi-perpendicular fast magnetosonic shock with wave precursor in collisionless plasma}

\begin{abstract}
A one-dimensional particle-in-cell (PIC) simulation tracks a fast magnetosonic shock over time scales comparable to an inverse ion gyrofrequency. The magnetic pressure is comparable to the thermal pressure upstream. The shock propagates across a uniform background magnetic field with a pressure that equals the thermal pressure upstream at the angle 85$^\circ$ at a speed that is 1.5 times the fast magnetosonic speed in the electromagnetic limit. Electrostatic contributions to the wave dispersion increase its phase speed at large wave numbers, which leads to a convex dispersion curve. A fast magnetosonic precursor forms ahead of the shock with a phase speed that exceeds the fast magnetosonic speed by about $\sim 30 \%$. The wave is slower than the shock and hence it is damped. 
\end{abstract}

\maketitle

Several particle-in-cell (PIC) simulation studies have found shocks that resemble their counterparts in a magnetohydrodynamic (MHD) plasma. The plasma model, on which PIC codes are based, assumes that effects caused by binary collisions between plasma particles are negligible compared to the collective interaction of the ensemble of plasma particles. We call such a plasma collisionless. Binary collisions are essential in an MHD model as they remove nonthermal plasma features and equilibrate the temperatures of all plasma species. 

Previous one-dimensional PIC simulations studied the propagation of MHD shocks across a perpendicular magnetic field. Shocks reached a steady state \cite{Forslund71,Dieckmann16,Dieckmann17} if they moved slow enough to avoid a self-reformation \cite{Guerolt17}. Self-reformation is a process that is not captured by an MHD model. If the shock propagates perpendicularly to the magnetic field then the dispersion relation of fast magnetosonic waves is concave for high frequency waves, which implies that their phase velocity decreases with increasing wave numbers; shock steepening drives slower waves that fall behind the shock as seen in Ref. \cite{Dieckmann17}. 

Here we demonstrate with a one-dimensional PIC simulation how turning the concave dispersion relation into a convex one removes the trailing wave and gives rise to a shock precursor. The precursor is formed by fast magnetosonic modes that outrun the shock. 

We compare aspects of the dispersion relation of a collisionless plasma with those of a single-fluid MHD model. The latter is valid for frequencies below the ion gyro-frequency $\omega_{ci}=ZeB_0/m_i$ ($Z,e,B_0,m_i$: ion charge state, elementary charge, amplitude of the background magnetic field and ion mass). One characteristic speed of this model is that of sound $\tilde{c}_s = {(\gamma p_0 / m_i n_0)}^{1/2}$, where $n_0$ is the plasma density, $p_0$ the thermal pressure and $\gamma=5/3$ the ratio of specific heats. The Alfv\'en speed $v_A = B_0 / {(\mu_0 m_in_0)}^{1/2}$ and $\tilde{\beta}=\tilde{c}_s^2/v_A^2$ equals the ratio of the plasma's thermal to magnetic pressure. 

The phase speed of waves in the MHD plasma depends on their propagation direction relative to the magnetic field. We define $\theta$ as the angle between the wave vector $\mathbf{k}$, which is parallel to the x-axis, and the magnetic field $\mathbf{B}_0=(B_0\cos{\theta},0,B_0\sin{\theta})$. Two waves exist if $\theta=0$; sound waves have the phase speed $\tilde{c}_s$ while that of the incompressible Alfv\' en waves is $v_A$. Only one propagating wave exists if $\theta = 90^\circ$: the fast magnetosonic mode with the phase speed $\tilde{v}_{fms}={(\tilde{c}_s^2+v_A^2)}^{1/2}$. 

Waves, which propagate obliquely to the magnetic field, can be subdivided into fast modes with the phase speed $v_f$ and slow modes with the phase speed $v_s$ with
\begin{equation}
\frac{2v_{f,s}^2}{v_A^2} = (1+\tilde{\beta}) \pm {\left ( {(1-\tilde{\beta})}^2 + 4\tilde{\beta}\sin^2{\theta} \right )}^{1/2}.
\label{MHDEqn}
\end{equation}
The fast mode (addition of both terms on right hand side in Eqn. \ref{MHDEqn}) is characterized by a magnetic pressure and a thermal pressure that are in phase while both pressures are in antiphase in the case of the slow mode \cite{Treumann11}. The phase speed of the slow mode goes to zero as $\theta \rightarrow 90^\circ$ and it becomes a tangential discontinuity. Magnetohydrodynamic shocks can be sustained by the slow and fast modes as well as by the sound wave \cite{Verscharen17}. 

A collisionless kinetic model describes each plasma species $L$ by a phase space density $f_L(\mathbf{x},\mathbf{v},t)$, from which the charge and current densities are obtained as $\rho_L = q_L\int f_L(\mathbf{x},\mathbf{v},t) \, d\mathbf{v}$ and $\mathbf{J}_L = q_L \int \mathbf{v} f_L(\mathbf{x},\mathbf{v},t) \, d\mathbf{v}$. Their summation over $L$ yields the total charge $\rho$ and current $\mathbf{J}$, which are coupled to the electric field $\mathbf{E}$ and the magnetic field $\mathbf{B}$ via Amp\`ere's law and Faraday's law. This model represents correctly the waves close to all resonances of a collisionless plasma. A PIC code approximates the phase space density distributions by computational particles (CPs) and their velocities are updated with the Lorentz force equation. The EPOCH code \cite{Arber15} we use fullfills Gauss' law and $\nabla \cdot \mathbf{B}=0$ exactly. 
    
The ion acoustic speed of electrons with the temperature $T_e$ and ions with the temperature $T_i$ in collisionless plasma is $c_s$ and the fast magnetosonic speed $v_{fms}={(c_s^2+v_A^2)}^{1/2}$ for $\theta = 90^\circ$. Both speeds are close to their MHD counterparts. Table \ref{table} defines further plasma parameters that determine the properties of a magnetized plasma. These are the plasma frequencies of the electrons $\omega_{pe}$ and ions $\omega_{pi}$ as well as the electron gyro-frequency $\omega_{ce}$. The electron thermal speed is $v_{the}$ and $r_{ge}$ is the electron's thermal gyroradius. The electron mass is $m_e$, $\epsilon_0$ is the vacuum permittivity and $\gamma_e=5/3$ and $\gamma_i = 3$ are the specific heat ratios for electrons and ions. 
\begin{table}[h]
\begin{tabular}{l r}
Parameter & Numerical value \\ 
$\omega_{pe} = (n_{e} e^{2}/\epsilon_{0} m_{e})^{1/2}$ & $9.35 \cdot 10^{11} s^{-1}$ \\ 
$\omega_{ce} = e B_{0}/ m_{e}$ & $1.5 \cdot 10^{11} s^{-1}$ \\
$v_{the} = (k_{B} T_{e}/m_{e})^{1/2}$ & $1.87 \cdot 10^{7} ms^{-1}$ \\
$r_{ge} = v_{the}/\omega_{ce}$ & $1.25 \cdot 10^{-4} m$ \\
$\omega_{pi} = (Z^{2} n_{i} e^{2}/\epsilon_{0} m_{i})^{1/2}$ & $1.54 \cdot 10^{10} s^{-1}$ \\ 
$\omega_{ci} = Z e B_{0}/ m_{i}$ & $4.07 \cdot 10^{7} s^{-1}$ \\
$\omega_{lh} = ((\omega_{ce} \omega_{ci})^{-1} +\omega_{pi}^{-2})$ & $2.46 \cdot 10^{9} s^{-1}$ \\
$c_{s} = ((\gamma_{e} T_{e} + \gamma_{i} T_{i})/m_{i})^{1/2}$ & $4.03 \cdot 10^{5} m/s$ \\ 
$v_{a} = B_{0}/(\mu_{0} n_{0} m_{i})^{1/2}$ & $7.9 \cdot 10^{5} m/s$  \\
$v_{fms} = (v_{a}^{2} + c_{s}^{2})^{1/2}$ & $8.88 \cdot 10^{5} m/s$  \\ 
\end{tabular} 
\caption{The plasma parameters in our simulation.}\label{table}
\end{table}

We use the following inital conditions for our simulation. We resolve one spatial dimension $x$ and three particle velocity components. Periodic boundary conditions are used for the fields and open boundary conditions for the computational particles (CPs). The simulation box is large enough to separate effects introduced by the boundaries from the area of interest. The length $L_0$ = 0.75 m of the simulation box is subdivided into evenly spaced grid cells with the length $\Delta_x = 5 \mu m$. We consider here fully ionized nitrogen. 

The ambient plasma fills the interval $0 < x < 2L_0 /3$. Its electron and ion temperatures are $T_e = 2.32 \times 10^{7} K$ and $T_i = T_e /12.5$. Table \ref{table} lists all relevant parameters of the ambient plasma with the ion density $n_i=n_0$ and the electron density $n_e=7n_0$ with $n_e=2.75\times 10^{20}m^{-3}$. A denser plasma fills the interval $-L_0 /3 \leq x \leq 0$. It consists of ions with the density $10n_0$ and the temperature $T_i$. The electrons have the density $70n_0$ and the temperature $3T_e$. All species are initially at rest. A spatially uniform background magnetic field with the strength $B_0 = 0.85$ T and orientation $\theta=85^\circ$ fills the entire simulation box. Our initial conditions match those in Ref. \cite{Dieckmann17} except for the magnetic field direction.

We represent the electrons and ions of the ambient plasma by $3 \times 10^{7}$ CPs each. Those of the dense plasma are each resolved by $4.5 \times 10^{7}$ CPs. The simulation box covers the interval $-2000 < x/r_{ge} < 4000$ ($r_{ge}:$ electron thermal gyroradius). We examine the data during the time interval $T_0 \leq t\omega_{ce} \leq T_{max}$ with $T_0 = 2 \times 10^4$ (130 ns) and $T_{max} = 2.4 \times 10^4$ (160 ns). $T_{max}$ is resolved by $9.52 \times 10^{6}$ time steps. 

\begin{figure}
\includegraphics[width=\columnwidth]{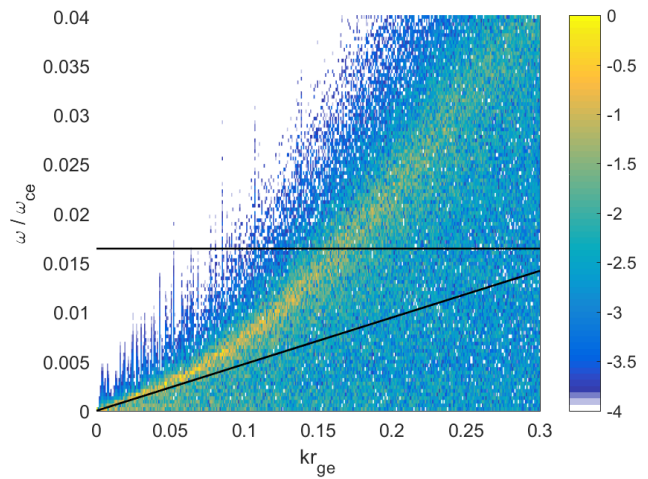}
\caption{The 10-logarithmic power spectrum $P_B(k,\omega)$ of $B_z(x,t)$. The dispersion relation $\omega_{f} = v_{f}v_{the}^{-1}k$ is overplotted and the horizontal line is $\omega=\omega_{lh}\omega_{ce}^{-1}$. Strong noise indicates weakly damped waves.}
\label{figure1}
\end{figure}

Equation \ref{MHDEqn} gives us the speeds $v_f \approx v_{fms}$ and $v_s \approx v_{fms}/25$ for $\theta = 85^\circ$ and  the dispersion relations of the slow and fast modes are $\omega_{s,f} = v_{s,f}k$. Their dispersion relation in the collisionless plasma can be estimated with a separate PIC simulation. It initializes a plasma with the parameters given in Table \ref{table} in a box with length 1 m and periodic boundary conditions and evolves the fields over the interval $0 \le t\omega_{ce}\le 1.7 \times 10^4$. Figure \ref{figure1} shows the power spectrum $P_B(k,\omega)$ of $B_z (x,t)$.

Strong noise indicates regions in $k,\omega$-space that are only weakly damped. The dispersion relation $\omega_f$ follows the frequency interval with strong noise for $kr_{ge}<0.05$. The frequency of the strong noise increases beyond $\omega_f$ for $kr_{ge}>0.05$. The dispersion relation is convex at such large $k$ \cite{Sagdeev66}. The band with the strong noise crosses $\omega_{lh}$, which is no longer a resonance for $\theta=85^\circ$, since $cos^{2} \theta \nleq m_e / m_i$ \cite{Verdon08}, and it gradually damps out with increasing $\omega$. Modes with $kr_{ge}\approx 0.15$ reach a frequency $\omega \approx 0.015\omega_{ce}$. Their phase speed is $v_{the}/10 \approx 2 v_f$. 

\begin{figure*}
\includegraphics[width=1.0\textwidth]{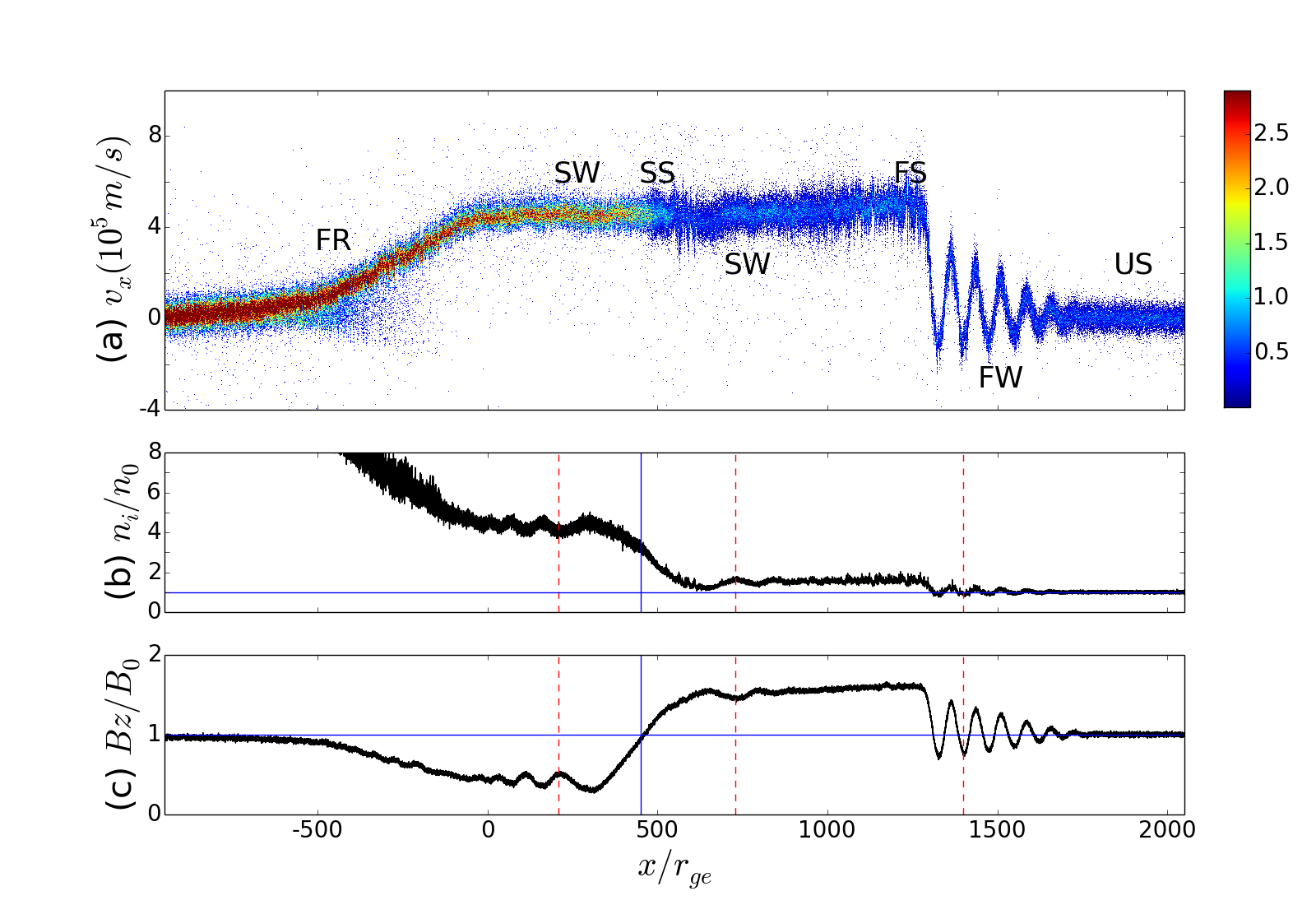}
\caption{The plasma state at the time $t\omega_{ce}=2 \times 10^4$: panel (a) shows the phase space density distribution of the ions normalized to the maximum upstream value and clamped at 2.9 for visualization reasons. We recognize the fast rarefaction wave (FR), the precursor wave (PW), the slow shock (SS), the fast shock (FS), the fast wave (FW) precursor and the upstream (US) ions. We observe slow mode waves close to the slow shock. Panel (b) shows the ion density $n_i/n_0$ . The blue lines denote $n_i=n_0$ and $x/r_{ge} = 450$. The magnetic $B_z$ component is plotted in (c). The blue line denotes $B_z = B_0$ and $x/r_{ge} = 450$. The dashed red lines in (b,c) emphasize the phase relation between $n_i$ and $B_z$ and, thus, the wave mode.}
\label{pxion_130ns_85deg_0}
\end{figure*}

Figure \ref{pxion_130ns_85deg_0} shows the ion phase space density, the ion density and the magnetic field at the time $T_0$. The ion's phase space density has its maximum at the left of Fig. \ref{pxion_130ns_85deg_0}(a) and the mean velocity of the ions vanishes. These are blast shell ions. The ions gain speed with increasing $x$ in the interval $-500 \le x/r_{ge} \le -50$ and their density decreases in Fig. \ref{pxion_130ns_85deg_0}(b). The acceleration is accomplished by the rarefaction wave that propagates to the left into the dense plasma and accelerates its ion to the right. The accelerated ions form a blast shell that expands with a constant speed and density up to $x/r_{ge} \approx 400$. The ion velocity remains constant but the density decreases from its value in the blast shell to the density $n_i \approx 1.5 n_0$. The magnetic field amplitude and, hence, its pressure increase as the ion density decreases and the anticorrelation of both is characteristic of a slow magnetosonic wave. A tangential discontinuity formed at this location in Ref. \cite{Dieckmann17}, which considered a magnetic field direction $\theta = 90^\circ$. The oblique magnetic field facilitates particle transport across the discontinuity, which changes the tangential discontinuity into a slow magnetosonic shock.

The source of the ions in the interval $600 \le x/r_{ge} \le 1350$ is the ambient plasma and they have been accelerated and compressed by the forward shock, which is located in Fig. \ref{pxion_130ns_85deg_0}(a) at $x/r_{ge}\approx 1350$. The ion density and the magnetic field amplitude both decrease with increasing $x$ across the shock and it is thus mediated by the fast magnetosonic mode. Strong waves, for which the ion density oscillates in phase with the magnetic amplitude, are observed between the shock and the upstream. Their amplitude of this shock precursor decreases with increasing $x$ and the phase relation between the thermal and magnetic pressure shows that it is formed by the fast magnetosonic mode. 
\begin{figure*}[htb]
 \subfloat[]{\includegraphics[width=0.34\textwidth]{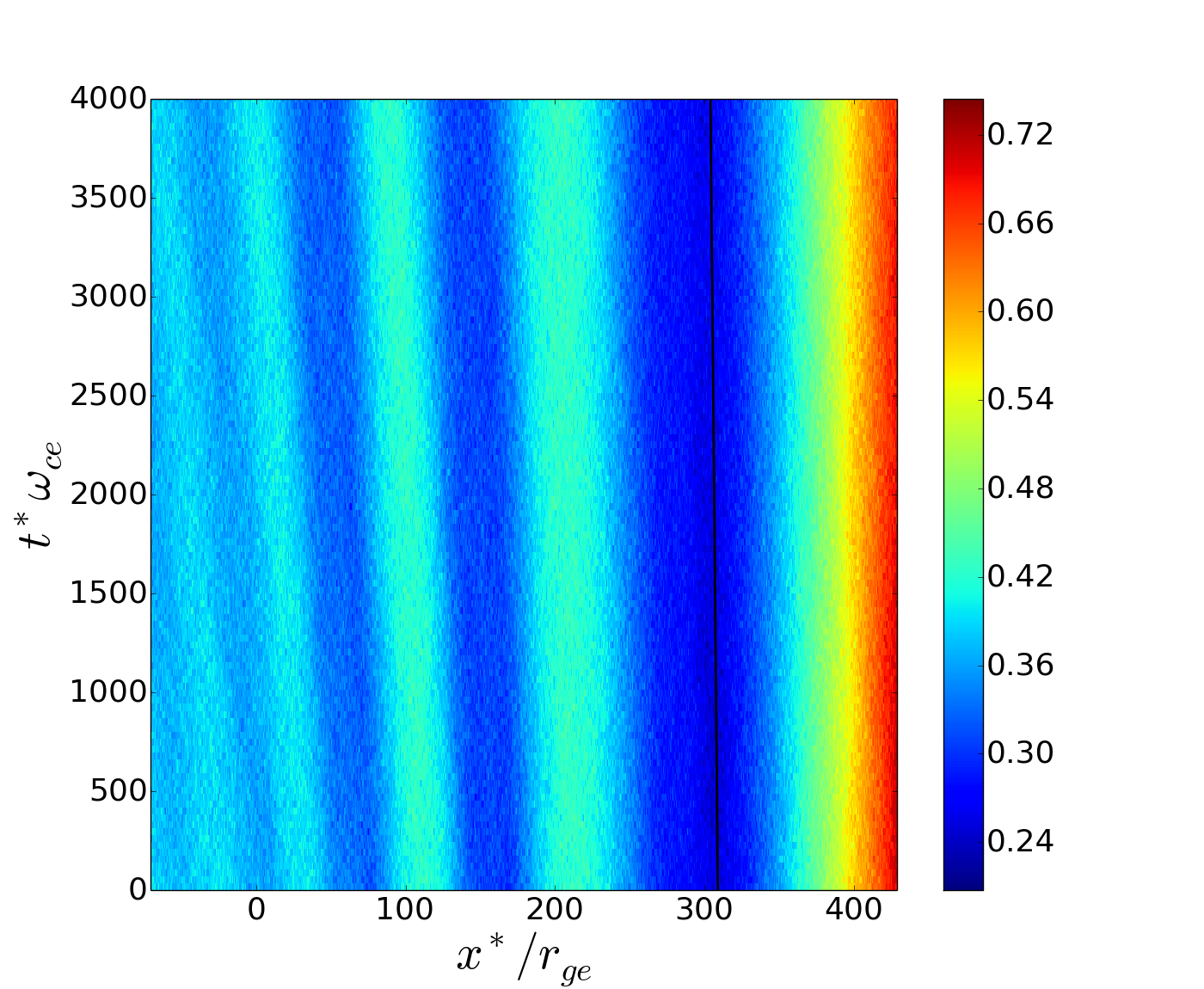}\label{im1}}
 \subfloat[]{\includegraphics[width=0.34\textwidth]{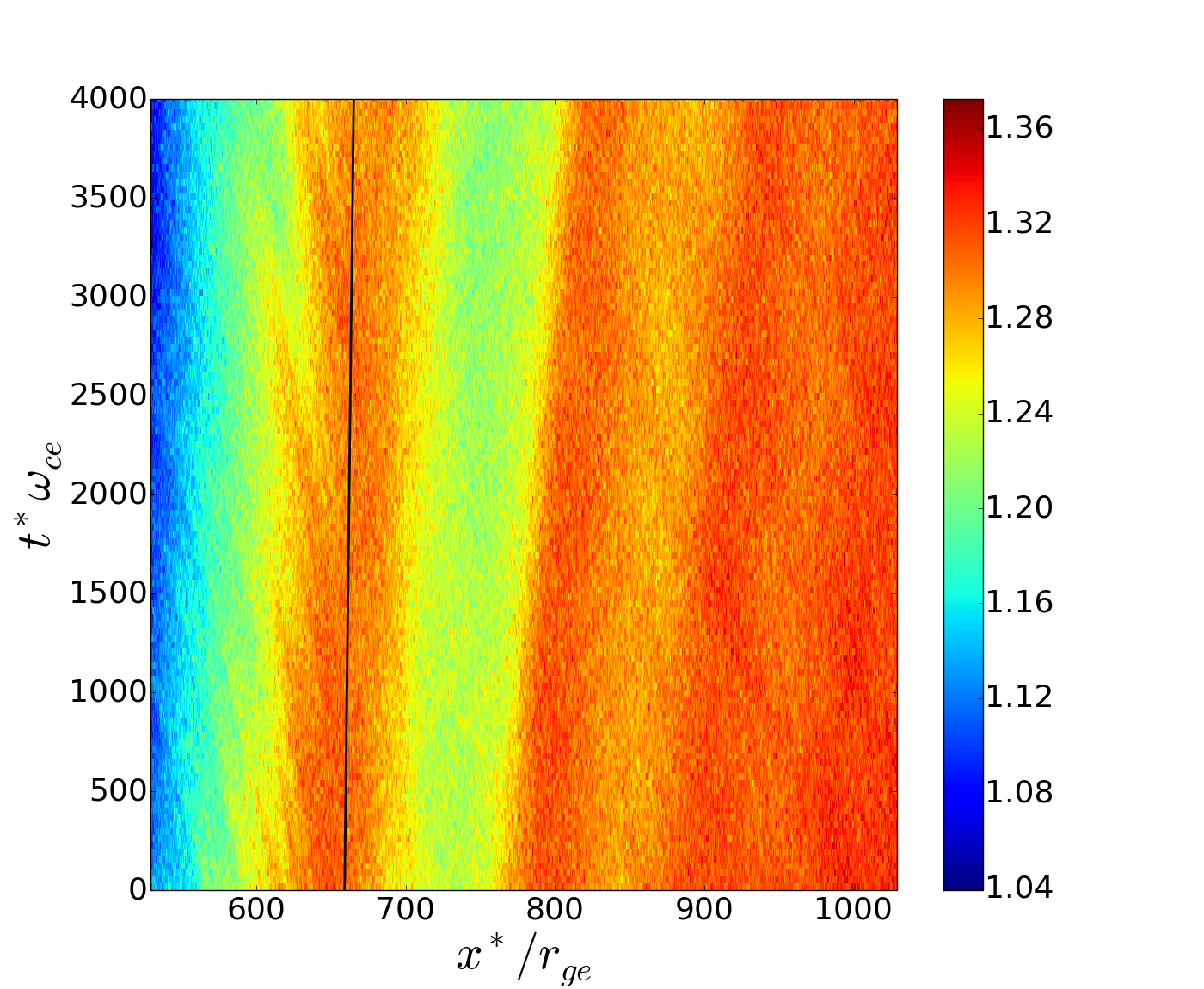}\label{im2}}
 \subfloat[]{\includegraphics[width=0.34\textwidth]{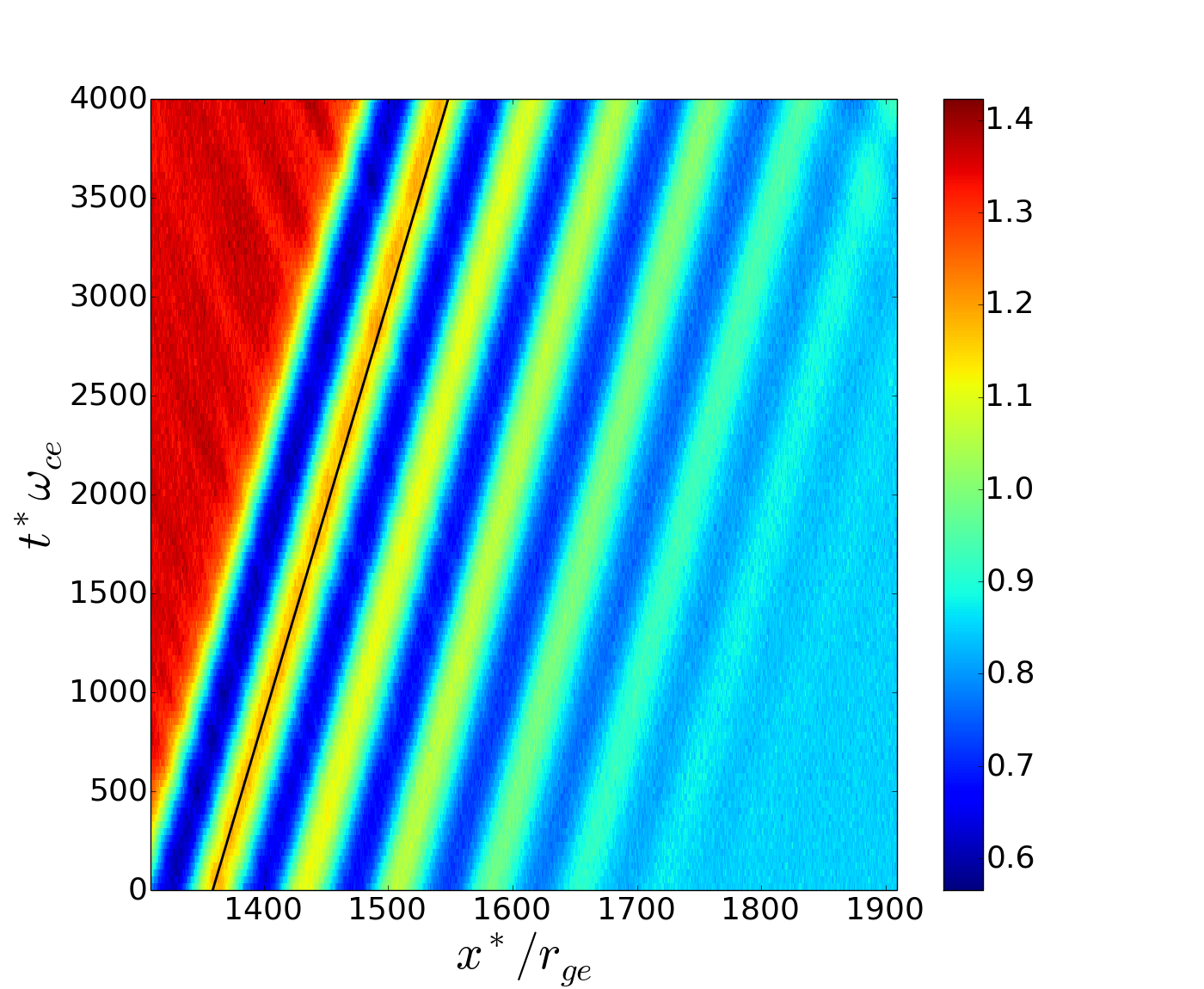}\label{im3}} 
 \caption{The evolution in time of $B_z$ in the reference frame that moves with $v_b$: panel (a) shows the field distribution of the magnetosonic waves to the left of the slow shock, panel (b) that of the waves to the right and panel (c) shows the magnetic field distribution of the fast shock and the precursor. The black lines in (a, b) have a slope that corresponds to the speed $v_s$, while that in panel (c) corresponds to the speed $v_f$.
\label{frame_tx_wkx_slow85deg} }
\end{figure*}
The precursor is a consequence of the convex dispersion relation observed in Fig. \ref{figure1}. Shock steepening drives waves with a large wave number, which outrun the shock and propagate upstream. The wavelength of the precursor waves is $\approx 80 r_{ge}$, which gives the wave number $kr_{ge}\approx 0.08$ and a phase speed $\approx 1.3 v_f$ (See Fig. \ref{figure1}). 

The precursor wave is damped with increasing $x>1350 r_g$.

Waves are also observed on both sides of the slow shock. The vertical dashed red lines demonstrate that the oscillations of the magnetic and thermal pressures have an opposite phase, which suggests that they are slow magnetosonic waves. We gain additional information about these waves and the precursor by examining their evolution in time in their rest frame. Figure \ref{pxion_130ns_85deg_0}(a) shows that the ions move at the spatially uniform mean speed $v_b \sim 4.5 \times 10^5 $ m/s. The simulation frame equals the upstream frame and $v_b$ is thus the speed of the rest frame of the waves in the upstream frame. We transform the distribution of $B_z(x,t)$ from the upstream frame into the rest frame of the waves for the times $0 \le t^* \omega_{ce} \le 4000$, where $t^*\omega_{ce} = t\omega_{ce} - 2 \times 10^4$. We transform space as $x^* = x-v_b t^*$ for the times $0 \le t^* \omega_{ce} \le 4000$. The moving frame matches that of Fig. \ref{pxion_130ns_85deg_0} at $t^*=0$.

Figure \ref{frame_tx_wkx_slow85deg} shows the wave fields in three intervals of the box. The wave fields close to the slow magnetosonic shock (location $x/r_{ge} \approx 400$) in Figs. \ref{frame_tx_wkx_slow85deg}(a, b) reveal waves that propagate away from the shock. Their phase speed $\sim v_s$ together with the phase relation between the magnetic pressure and the thermal pressure in Fig. \ref{pxion_130ns_85deg_0} demonstrates that these are slow magnetosonic waves. The shock in Fig. \ref{frame_tx_wkx_slow85deg}(c) propagates at the speed $v_f$ in the downstream frame of reference and at $v_f + v_b \simeq 1.5 v_{fms}$ in the upstream frame. The precursor waves outrun the shock but their phase speed $\sim 1.3 v_f$ imples that they are too slow to be undamped modes. 

In summary we have modeled the expansion of a dense plasma into a dilute ambient one in the presence of an initially spatially uniform quasi-perpendicular magnetic field. The thermal pressure jump between the dense and dilute plasma drove a fast mode rarefaction wave, which propagated into the dense plasma and launched a blast shell into the ambient plasma. A slow mode shock formed at the boundary between the blast shell plasma and the shocked ambient plasma. The shocked ambient plasma was separated from the pristine ambient plasma by a fast magnetosonic shock. The convex dispersion relation of the fast magnetosonic modes gave rise to a shock precursor. The precursor softened the transition of the ambient plasma into the shocked ambient one and we could not observe a strong acceleration of ions by the shock passage. An absent shock-reflected ion beam implied that the quasi-perpendicular shock did not reform by driving solitons upstream \cite{Guerolt17}.

\textbf{Acknowledgements:} the simulations were performed on resources provided by the Grand Equipement National de Calcul Intensif (GENCI) through grants ... The EPOCH code has been developed with support from EPSRC (grant No: EP/P02212X/1).
This work was supported by the French National Research Agency Grant ANR-14-CE33-0019 MACH. 
This work was also granted access to the HPC resources of CINES and TGCC under allocations A0020510052 and A0030506129 made by GENCI (Grand Equipement National de Calcul Intensif), and has been partially supported by the 2015-2019 grant of the Institut Universitaire de France.
This study has been carried out with financial support from the French State, managed by the French National Research Agency (ANR) in the frame of
``the Investments for the  future''  Programme IdEx Bordeaux-ANR-10-IDEX-03-02.


\begin{thebibliography}{}
\bibitem{Forslund71} D. W. Forslund, and J. P. Freidberg, Phys. Rev. Lett. \textbf{27}, 1189 (1971).
\bibitem{Dieckmann16} M. E. Dieckmann, G. Sarri, D. Doria, A. Ynnerman, and M. Borghesi, Phys. Plasmas \textbf{23}, 062111 (2016).
\bibitem{Dieckmann17} M. E. Dieckmann, D. Folini, R. Walder, L. Romagnani, E. d'Humieres, A. Bret, T. Karlsson, and A. Ynnerman, Phys. Plasmas \textbf{24}, 094502 (2017).
\bibitem{Guerolt17} R. Guerolt, Y. Ohsawa, and N. J. Fisch, Phys. Rev. Lett. \textbf{118}, 125101 (2017).
\bibitem{Treumann11} A. Balogh, R. A. Treumann, ISSI Scientific Report Series \textbf{12}, (2011)
\bibitem{Verscharen17} D. Verscharen, C. H. K. Chen, and R. T. Wicks, Astrophys. J. \textbf{840}, 106 (2017).
\bibitem{Arber15} T. D. Arber, K. Bennett, C. S. Brady, A. Lawrence-Douglas, M. G. Ramsay, N. J. Sircombe, P. Gillies, R. G. Evans, H. Schmitz, A. R. Bell, and C. P. Ridgers, Plasma Phys. Controll. Fusion \textbf{57}, 113001 (2015).
\bibitem{Sagdeev66} R.Z. Sagdeev, Rev. Plasma Phys. \textbf{4}, (1966)
\bibitem{Verdon08}  A. Verdon, I. Cairns, D. Melrose, and P. Robinson, Proceedings of the International Astronomical Union, 4(S257), 569-573. (2008).
\end{thebibliography}
\end{document}